
\input phyzzx
\normalbaselineskip=30 pt plus 0.3 pt minus 0.15 pt
\baselineskip=\normalbaselineskip

\pubnum={03}

\date{February 1993}
\titlepage
\title {Gravitational form factors of the neutrino}
\author{K. L. Ng*}
\footnote*{Present address: Institute of Physics, Academia Sinica, Nankang,
Taipei 115, Taiwan.}

\address{Department of Physics, National Taiwan University, Taipei 115, Taiwan}

\def\ab{{\alpha\beta}}
\def\em{{energy-momentum }}

\def\Gab{{\Gamma_{\alpha\beta}}}

\def\uf{{\overline u_f}}
\def\ui{{u_i}}

\def\ampfi{{<\nu_f(p_f)|\theta_{\alpha\beta}|\nu_i(p_i)>}}
\def\ampif{{<\nu_i(p_i)|\theta_{\alpha\beta}|\nu_f(p_f)>}}
\def\ampii{{<\nu_f(p_i')|\theta_{\alpha\beta}|\nu_i(p_i)>}}

\def\Mampfi{{<\nu^M _f(p_f)|\theta _\ab|\nu^M _i(p_i)>}}
\def\Mampif{{<\nu^M _i(p_i)|\theta _\ab|\nu^M _f(p_f)>}}

\def\g5{{\gamma_5}}
\def\dfi{{\delta m_{fi}^2}}
\def\dmfi{{\delta m_{fi}}}
\abstract
The gravitational properties of the neutrino is studied
in the weak field
approximation.  By imposing the hermiticity condition, CPT and CP invariance
on the \em tensor matrix element, we shown that the
allowed gravitational
form factors for Dirac and Majorana neutrinos are very different.
In a CPT and CP invariant theory, the \em tensor for a Dirac neutrino of
the same specie is characterized by
four gravitational form factors.  As a result of CPT invariance,
the \em tensor for a Majorana neutrino of the same specie
has five form factors.
In a CP invariant theory, if the initial and final Majorana
neutrinos have the same (opposite) CP parity, then only tensor (pseudo-tensor)
type transition are allowed.


\endpage

{\bf \centerline{1.  Introduction}}

\par
If a neutrino has mass, then the question of whether the neutrino is a Dirac or
Majorana type particle arise naturally.  This is because the neutrino may be
its own anti-particle (Majorana particle).  The difference between a Dirac and
Majorana neutrino is clearly exhibited in the neutral current interaction
process [1],
observation of neutrinoless double beta decay, and in their electromagnetic
properties [2,3].  For example, a spin 1/2 Majorana neutrino can only have the
anapole moment form factor if CPT invariance holds.  This result was
generalized to an arbitrary half integral spin Majorana fermion in Ref. [4],
and an arbitrary spin Majorana fermion in Ref. [5].
\par
However, there are relatively few discussions on the
gravitational properties of a spin 1/2 fermion [6, 7].
In this paper, we extend their work by performing a complete study of the
gravitational properties of the neutrino.
In section 2, we present a general analysis of the \em tensor
$\theta_{\alpha\beta}$ matrix element between two spin 1/2 neutrinos.
Using the Dirac equation, the symmetric properties of $\theta_{\alpha\beta}$
and the \em conservation condition, we arrive at the most general
expression for the gravitational form factors of the neutrino.
By imposing the hermiticity condition, CPT and
CP invariance on the $\theta_{\alpha\beta}$
matrix element, we obtained certain conditions on the gravitational
form factors for the neutrino.  We summarize the results in section 3.

\vskip 1 true cm

{\bf\centerline{2. Gravitational form factors of the neutrino}}

\par
In this section we study the allowed form of the couplings for the
energy-momentum tensor $\theta_\ab$ matrix element between two neutrino
states.  We carry out the analysis in the weak field approximation,
$$g_\ab = \eta_\ab + \kappa h_\ab    \eqno(2.1)$$
where $\eta_\ab$ is the flat space-time metric, $h_\ab$ is the graviton field,
and $\kappa=32\pi G$.
In our paper we closely follow the notations used in Ref.[3].

\vskip 1 true cm

{\bf\centerline{A. General analysis}}
\par

Consider the invariant amplitude for the process $\nu_i \rightarrow \nu_f + g$,
where $\nu_i$ and $\nu_j$
are two Dirac neutrinos with masses $m_i$ and $m_j$ $(m_i>m_f)$ and $g$ is
the graviton (virtual or real).  The transition amplitude for this process
is given by
$$\ampfi = \overline u(p_f) (\Gamma _\ab)_{fi} u(p_i)   \eqno(2.2)$$
where $|\nu_i>$ and $<\nu_f|$ are the initial and final neutrino states
respectively, and $(\Gamma _\ab)_{fi}$ is the dressed vertex function that
characterizes the above invariant amplitude.

Lorentz invariance implies that the vertex function in general can
have twenty-four types of coupling: twelve tensor types and twelve
pseudo-tensor types. The twelve possible tensor types of
coupling have the following forms:
$\eta_\ab$, $q_\alpha q_\beta$, $\{qP\}_\ab$,
$\{q \gamma\}_\ab$,
$P_\alpha P_\beta$,
$\{P \gamma\}_\ab$,
$\{\sigma _{\alpha \mu} q^\mu q_\beta \}_\ab = \{\sigma q q\}_\ab$,
$\{\sigma q P\}_\ab$,
$\{\sigma q \gamma\}_\ab$,
$\{\sigma P P\}_\ab$,
$\{\sigma P q\}_\ab$  and
$\{\sigma P \gamma\}_\ab$, where we have suppressed the Lorentz indicies,
$\{ \}_\ab$ denote symmetrization over the indices $\alpha$ and $\beta$,
$q=p_f-p_i$, $P=p_f+p_i$ and
$\sigma=\sigma _{\alpha\mu}={i \over 2} [\gamma _\alpha , \gamma _\mu]$.
The pseudo-tensor types of coupling are obtained
by the addition of a $\gamma_5$ factor.

Using the Dirac equation, $(\gamma_\mu p^\mu - m)u = 0$, one obtains
identities which relate the various
types of coupling (like the Gordon decomposition relation), and hence reduces
the number of independent couplings.  We collect these
relations in the appendix.  Thus, the \em matrix element
between two Dirac neutrino states may be written as
$$
\eqalign{ <\nu_f(p_f)|&\theta_{\alpha\beta}|\nu_i(p_i)>
= \overline u(p_f) (\Gamma _\ab)_{fi} u(p_i) \cr
&= \overline u(p_f)
\bigg [
 T_{1fi} \eta_\ab + {T_{2fi}} q_\alpha q_\beta + T_{3fi} P_\alpha P_\beta \cr
     & + T_{4fi}  \{q P\}_\ab + T_{5fi} \{\sigma q q \} _\ab
       + T_{6fi} \{\sigma q P\}_\ab \cr
      &+ P_{1fi} \g5 \eta_\ab + {P_{2fi}} \g5 q_\alpha q_\beta
       + P_{3fi} \g5 \{q \gamma \}_\ab \cr
     &+ P_{4fi}  \g5  \{P \gamma \}_\ab  + P_{5fi} \g5 \{\sigma q q \}_\ab
        +P_{6fi} \g5 \{\sigma q P\}_\ab  \bigg ]  u(p_i) \cr }   \eqno(2.3)$$
where $T=T(q^2,m_i,m_f)$ and $P=P(q^2,m_i,m_f)$ are the tensor and
pseudo-tensor types form factors respectively.

Conservation of the \em tensor ($q^\beta \theta_\ab =0$) implies the following
relations among the form factors,
$$   T_1 + T_2 q^2 + \dfi T_4 = 0	        \eqno(2.4)          $$
$$   \delta m_{fi}^2  T_3 +  q^2 T_4 =0	        \eqno(2.5)          $$
$$   q^2 T_5 + \dfi T_6 =0	                \eqno(2.6)          $$
$$   P_1 + P_2 q^2 - M_{if} P_3 =0	        \eqno(2.7)          $$
$$   i(q^2 P_5 + \delta m_{fi}^2 P_6) - M_{if} P_4 =0	\eqno(2.8)  $$
$$   q^2 P_3 + 2 \dfi P_4 =0	                \eqno(2.9)          $$
where $M_{if}=m_i + m_f$, and $\dfi = m_f ^2 - m_i ^2$.
Lorentz invariance and \em conservation imply that for $q^2 \not= 0$,
the general form for the \em martix element between two neutrinos states
is given by
$$
\eqalign{ <\nu_f(p_f)|&\theta_{\alpha\beta}|\nu_i(p_i)>
= \overline u(p_f)
\bigg [
       T_{1fi}         (\eta_\ab -         {1\over \dfi } \{qP\}_\ab
                              + {q^2 \over {(\dfi)^2} } P_\alpha P_\beta) \cr
+& {T_{2fi}}\bigg (q_\alpha q_\beta - { q^2\over \dfi } \{qP\}_\ab
                         + {q^4 \over {(\dfi)^2} } P_\alpha P_\beta \bigg) \cr
+&T_{6fi}\bigg (\{\sigma qP\}_\ab - {\dfi \over q^2} \{\sigma qq\}_\ab\bigg)
\cr
+& P_{1fi}              \g5 \bigg(\eta_\ab + {1 \over{M_{if}}} \{q\gamma\}_\ab
                        -{ q^2 \over{2 \dfi M_{if} } } \{P\gamma\}_\ab
                        +{ i \over{2 \dfi} } \{\sigma qq\}_\ab  \bigg )
\cr
+& {P_{2fi}}\g5 \bigg(q_\alpha q_\beta +{q^2 \over{M_{if}}}\{q\gamma\}_\ab
                        -{ q^4 \over{2 \dfi M_{if} } } \{P\gamma\}_\ab
                        +{ i q^2\over{2 \dfi} } \{\sigma qq\}_\ab  \bigg)   \cr
+& P_{6fi}\g5 \bigg(\{\sigma qP\}_\ab - {\dfi \over q^2} \{\sigma
qq\}_\ab\bigg)\bigg ] u(p_i) \cr }
\eqno(2.10)
$$
For the same neutrino flavor $m_i=m_f$, the solutions for eq.(2.4) to eq.(2.9)
are $T_4=T_5=0$, $T_2=-{T_1 \over q^2}$, $P_3=0$,
$P_2=-{P_1 \over q^2}$ and $P_5=-{2miP_4 \over q^2}$.

Thus the \em matrix element is reduced to
$$
\eqalign{ \ampii =&  \overline u(p_i')
\bigg [
 T_{1ii}(\eta_\ab -{{q_\alpha q_\beta}\over q^2}) +T_{3ii} P_\alpha P_\beta \cr
+&T_{6ii} \{\sigma q P \}_\ab
+ P_{1ii} \g5 (\eta_\ab -{{q_\alpha q_\beta}\over q^2}) \cr
+&P_{4ii} \g5 \{(\gamma -{\gamma _\mu q^\mu q \over q^2})P \}_\ab
+ P_{6ii} \g5 \{\sigma q P \}_\ab
\bigg ] u(p_i) \cr}                                            \eqno(2.11)
$$
This result agrees with Ref. [6, 7] except the $P_4$ term.
In analogy to the electromagnetic form
factors, $T_6$ is called the anomalous gravitational magnetic moment form
factor, $P_4$ the gravitational anapole moment form factor, and $P_6$
the gravitational dipole moment form factor [6].

\vskip 1 true cm
{\bf\centerline{B. Gravitational form factors of a Dirac neutrino}}

The \em tensor $\theta_\ab$ is proportional to
$p_\alpha p_\beta$ [8], where $p_\alpha = (ip_0,{\bf{p}}$), whereas the
hermiticity of the \em tensor
operator is given by,
$\theta_\ab ^\dagger = \eta_\alpha \eta_\beta \theta_\ab$.  The hermiticity
condition implies
$$
\ampfi ^\dagger =  \eta_\alpha \eta_\beta \ampif      \eqno(2.12)
$$
where $\eta_\alpha=(-1,1,1,1)$.
As a result of hermiticity, we have
$$     \gamma_0 (\Gab)^{\dagger}_{fi} \gamma_0
      = \eta_\alpha \eta_\beta (\Gab)_{if}.                  \eqno(2.13)  $$
This implies the following relations among the gravitational form factors,
$$     (T_1, T_2, T_6, P_1, P_2, P_6)^* _{fi} =
       (T_1, T_2, -T_6, -P_1, -P_2, P_6) _{if}   \eqno(2.14)             $$
and
$$      (T_1, T_3, T_6, P_1, P_4, P_6)^* _{ii} =
        (T_1, T_3, -T_6, -P_1, P_4, P_6) _{ii}.   \eqno(2.15)           $$
For the off-diagonal case, $\nu_f \not= \nu_i$, hermiticity does not
put any restriction on the form factors.  For the diagonal case, hermiticity
requires that all the form factor are real except for $T_6$ and $P_1$.

Under the CPT transformation,
$\theta_\ab \buildrel CPT \over \longrightarrow \theta _\ab$ and
$$
_{CPT} \ampfi _{CPT} = \ampif.    	\eqno(2.16)
$$
In terms of the Dirac spinor, the left-handed side of eq.(2.16) can be written
as
$$  _{CPT} \ampfi _{CPT}
= \overline u_{CPT}(-p_i)(\overline \Gamma_\ab)_{fi} u_{CPT}(-p_f)
\eqno(2.17)$$
and $u_{CPT}(p)$ is the CPT conjugate of the spinor $u(p)$,
$$ u_{CPT}(-p) = \gamma_0 V_{T}^{-1}(C \overline u ^t(p))^* 	\eqno(2.18)  $$
where $V_T$ is the
time-reversal matrix, t denotes the transpose operation, and
$\overline \Gamma_\ab$ is the vertex function describing the process
$\overline \nu_i \rightarrow \overline \nu_f + g$, where $\overline \nu$
denotes the anti-neutrino state.

Using Eq.(2.17) and the transformation properties of the gamma matrices under
the operators C and $V_T$
in Eq.(2.16), we obtain
$$
CV_T (\overline \Gamma _\ab)_{fi} V^{-1}_T C^{-1} = - (\Gamma _\ab)_{if}
                                          \eqno(2.19)
$$
As a result of CPT invariance, we obtained the following relations among the
form factors,
$$
(\overline T_1, \overline T_2,\overline T_6,
 \overline P_1, \overline P_2, \overline P_6)_{fi}
=
(-T_1, -T_2, T_6, -P_1, -P_2, P_6)_{if}        \eqno(2.20)
$$
and
$$
(\overline T_1, \overline T_3,\overline T_6,
 \overline P_1, \overline P_4, \overline P_6)_{ii}
=
(-T_1, -T_3, T_6, -P_1, P_4, P_6)_{ii}.      \eqno(2.21)
$$

\par
Under the CP transformation,
$\theta_\ab \buildrel CP \over \longrightarrow \eta_\alpha \eta_\beta
\theta_\ab$,
$$
_{CP} \ampfi _{CP} =  \eta_\alpha \eta_\beta \ampfi.      \eqno(2.22)
$$
The left-handed side of eq.(2.22) is given by
$$
_{CP} \ampfi _{CP}
=\overline u_{CP}(-p'_i) (\overline \Gamma'_\ab)_{fi} u_{CP}(-p'_f) \eqno(2.23)
$$
where $p'_\alpha = -\eta_\alpha p_\alpha = (ip_0,-{\bf{p}})$,
$\overline \Gamma '$ denotes the dressed vertex function
with $q$ and $P$ replaced by $q'$ and $P'$ and
$$
u_{CP}(-p')=\gamma_0 C \overline u ^t(p).		\eqno(2.24)
$$
Inserting Eq. (2.23) and Eq. (2.24) into Eq. (2.22), we obtain
$$
\gamma _0 C (\overline \Gamma ' _\ab)_{fi}^t C^{-1} \gamma_0
= - \eta_\alpha \eta_\beta (\Gamma _\ab)_{fi}.		\eqno(2.25)
$$
If CP invariance holds, we obtain the following relations among
the form factors,

$$
(\overline T_1, \overline T_2,\overline T_6,
 \overline P_1, \overline P_2, \overline P_6)_{fi}
=
(-T_1, -T_2, T_6, P_1, P_2, -P_6)_{fi}        \eqno(2.26)
$$
and
$$
(\overline T_1, \overline T_3,\overline T_6,
 \overline P_1, \overline P_4, \overline P_6)_{ii}
=
(-T_1, -T_3, T_6, P_1, P_4, -P_6)_{ii}.       \eqno(2.27)
$$
For the diagonal case, it follows from the CPT and CP invariance that
$P_{1ii}=P_{6ii}=0$.  That means
in a CPT invariant theory, a Dirac neutrino cannot have the form factors $P_1$
and $P_6$ if the interaction respects CP symmetry.
\endpage
{\bf\centerline{C. Gravitational form factors of a Majorana neutrino}}

Under the CPT transformation a Majorana neutrino $\nu^M$ transforms as [9]
$$
CPT|\nu^M({\bf{p}},s)> = \eta _{CPT}^s |\nu ^M ({\bf{p}},-s)>	\eqno(2.28)
$$
where $\eta _{CPT}^s$ is a phase factor that depends on the spin of the
particle, with $\eta _{CPT}^s = -\eta _{CPT}^{-s}$.  Assuming CPT invariance
for the \em tensor matrix element, we have
$$
_{CPT} \Mampfi _{CPT} = \Mampif				\eqno(2.29)
$$
For a Majorana neutrino, the left-hand side of Eq. (2.29) can be
written as
$$
_{CPT} \Mampfi _{CPT}
= \overline u_{PT}(p_f) (\Gamma_\ab)_{fi} u_{PT}(p_i)	\eqno(2.30)
$$
where
$u_{PT}(p) = \gamma _0 V^{-1}_T u^*(p)$.  This implies that
$$
V_T^{-1} (\Gamma _\ab^t)_{fi} V_T = (\Gamma _\ab)_{if}.	\eqno(2.31)
$$
Using the transformation properties of the gamma matrices under the operator
$V_T$ in Eq.(2.31), then as a
result of CPT invariance, we obtain the following relations among the
form factors,

$$
(T_1, T_2, T_6, P_1, P_2, P_6)_{fi} =
(T_1, T_2, T_6, P_1, P_2, P_6) _{if}   \eqno(2.32)
$$
and
$$
(T_1, T_3, T_6, P_1, P_4, P_6) _{ii} =
(T_1, T_3, T_6, P_1, -P_4, P_6) _{ii}.  \eqno(2.33)
$$

For the same neutrino species, CPT invariance implies that $P_4=0$,
that is a Majorana neutrino cannot have the gravitational anapole moment form
factor.

\par

Under CP transformation, a Majorana neutrino transforms as
$$
CP|\nu^M({\bf{p}},s)> = \eta _{CP}^* |\nu ^M (-{\bf{p}},s)>	\eqno(2.34)
$$
where $\eta_{CP}^*$ is the CP parity of the Majorana neutrino with
$\eta_{CP}^*= \pm i$.
Assuming CP invariance we have
$$
_{CP} \Mampfi _{CP}
= \eta_\alpha \eta_\beta \Mampfi.			\eqno(2.35)
$$
The left-hand side of Eq. (2.35) can be
written as
$$
_{CP} \Mampfi _{CP}
= \overline u_{P}(p'_f) (\Gamma'_\ab)_{fi} u_{P}(p'_i)	\eqno(2.36)
$$
where
$u_{P}(p') = \gamma_0 u(p)$.
Using Eq.(2.35) and Eq.(2.36), we obtain
$$
\eta^i \eta^f \gamma_0 (\Gamma ' _\ab)_{fi} \gamma _0
= \eta_\alpha \eta_\beta (\Gamma _\ab)_{fi}		\eqno(2.37)
$$
where $\eta_{CP}= i \eta$. As a result of CP invariance, we obtain the
following relations among the form factors,

$$
\eta^i \eta^f (T_1, T_2, T_6, P_1, P_2, P_6)_{fi}
= (T_1, T_2, T_6, -P_1, -P_2, -P_6)_{fi}                \eqno(2.38)
$$
and
$$
(T_1, T_3, T_6, P_1, P_4, P_6)_{ii}
= (T_1, T_3, T_6, -P_1, -P_4, -P_6)_{ii}.                 \eqno(2.39)
$$

We observe that the amplitude for the process
$\nu^M_i \rightarrow \nu^M_f + g$ depends on the relative CP parity of the
initial and final neutrino states.  For instance,
if $\eta^i \eta^f =1$, a Majorana neutrino has a tensorial type of transition
form factors, while for $\eta^i \eta^f = -1$, a Majorana neutrino has a
pseudo-tensorial type of transition form factors.
\vskip 1 true cm
{\bf{\centerline{3. Summary}}}
It is shown that the invariant amplitude for the process
$\nu_i \rightarrow \nu_f + g$ is characterized by six gravitational form
factors (three tensor and three pseudo-tensor types).
The hermiticity condition requires that four of the form factors are real.
As a result of CPT and CP invariance, a Dirac neutrino of the same species has
four gravitational form factors. A Majorana neutrino has five form factors
(no gravitational anapole moment form factor) as a result of
CPT invariance, which is agree with Ref. [7] result.
In a CP invariant theory, if the initial and final Majorana
neutrinos have the same (opposite) CP parity, then only tensor (pseudo-tensor)
type transition are allowed.   For the same
neutrino species, the \em matrix element for a Majorana neutrino
is characterized by tensor couplings only [7].

\vskip 1 true cm
{\bf{\centerline{Acknowledgments}}}
This work is supported by the National Science Council of the R.O.C
research grant NSC-81-0208-M-002-518.

\endpage
\vskip 3 true cm
{\bf{\centerline{Appendix}}}
In this appendix we present the identities among the various types of
coupling that were employed in our calculation.

{\bf{Tensor type couplings}}
$$
\uf \{\sigma q q \} \ui = i\uf \{q P- M_{if} q \gamma\} \ui
							\eqno(A.1)
$$
$$
\uf \{\sigma q P \}  \ui= i\uf \{PP - M_{if} P \gamma \} \ui
\eqno(A.2)
$$
$$
\uf \{\sigma q \gamma \}\ui =i\uf(\{q \gamma \}- 2\dmfi \eta_\ab) \ui
\eqno(A.3)
$$
$$
\uf \{\sigma P P \} \ui = i\uf \{qP - \dmfi P \gamma \} \ui         \eqno(A.4)
$$
$$
\uf \{\sigma Pq\} \ui = i\uf (qq - \{ \dmfi q \gamma \}) \ui         \eqno(A.5)
$$
$$
\uf \{\sigma P \gamma \} = i\uf (\{P \gamma \} - 2M_{if}\eta_\ab) \ui
\eqno(A.6)
$$
{\bf{Pesudo-tensor type couplings}}
$$
\uf \g5 \{\sigma q q \} \ui = i\uf \g5 \{Pq + \dmfi q \gamma \} \ui
\eqno(A.7)
$$
$$
\uf \g5 \{\sigma q P \}  \ui= i\uf \g5 \{PP + \dmfi P \gamma \} \ui
\eqno(A.8)
$$
$$
\uf \g5 \{\sigma q \gamma \}\ui =i\uf \g5 (\{q \gamma \} + 2M_{if}\eta_\ab)\ui
\eqno(A.9)
$$
$$
\uf \g5 \{\sigma P P \} \ui = i\uf \g5 \{qP + M_{if} P \gamma \} \ui
\eqno(A.10)
$$
$$
\uf \g5 \{\sigma Pq\} \ui =i\uf \g5 (qq + \{ M_{if} q\gamma \} ) \ui
\eqno(A.11)
$$
$$
\uf \g5 \{\sigma P \gamma \} = i\uf \g5 (\{P \gamma \} + 2 \dmfi \eta_\ab) \ui
\eqno(A.12).
$$
where $\dmfi=m_f-m_i$ and $M_{if}=m_i+m_f$.

\ref{B. Kayser and R.E. Shrock, Phys. Lett. {\bf{112B}}, 137 (1982).}

\ref{    B. Kayser, Phys. Rev. D, {\bf{26}}, 1662 (1982); J. F. Nieves,
Phys. Rev. D {\bf{26}}, 3152 (1982).}

\ref{S.M. Bilenky and S.T. Petcov, Rev. Mod. Phys. {\bf{59}}, 671 (1987).}

\ref{    E.E. Radescu, Phys. Rev. D {\bf{32}}, 1266 (1985).}

\ref{   F. Boudjema, C. Hamzaoui, V. Rahel and H.C. Ren, Phys. Rev. Lett
{\bf{62}}, 852 (1989).}

\ref{    I. Yu. Kobzarev and L.B. Okun, Sov. Phys. JETP {\bf{16}}, 1343
(1963).}

\ref{     A. Khare and J. Oliensis, Phys. Rev. D {\bf{29}}, 1542 (1984).}

\ref{ The \em tensor $\theta_{\ab}$ is proportional to $p_\alpha p_\beta$,
where
$p$ can be either $p_i$ or $p_f$.  See S. Weinberg, {\it{Gravitation and
Cosmology}}, John Wiley \& Sons, New York 1972}

\ref{    B. Kayser and A. Goldhaber, Phys. Rev. D {\bf{28}}, 2341 (1983)}
\referencecount=-1
\chapterskip=0pt

\refout
\endpage

\bye